\documentclass[twocolumn,aps,pra]{revtex4-2}
\usepackage{bm,bbm}
\usepackage{amsfonts,amsmath,amssymb,latexsym}
\usepackage{graphicx}
\usepackage{mathtools}
\usepackage{xcolor}
\usepackage{hyperref}
\begin{document} 
\title{Guiding polarizable particles in multi-hole Gaussian beams}
\author{Tomasz Rado\.zycki}
\email{t.radozycki@uksw.edu.pl}
\affiliation{Faculty of Mathematics and Natural Sciences, College of Sciences, Institute of Physical Sciences, Cardinal Stefan Wyszy\'nski University, W\'oycickiego 1/3, 01-938 Warsaw, Poland} 

\begin{abstract}
The present paper discusses the application of certain special Gaussian beams that, thanks to some polynomial prefactors, have uniquely designed holes in the irradiance. Such holes, or rather tubes, can constitute potential valleys for negatively polarizable particles, providing the possibility of guiding several objects of that kind, each along its own trajectory. 
The mechanism of creating these holes by interference of Gaussian beams which exhibit orbital angular momentum is discussed, and then the trajectories of particles moving in such a wave are numerically calculated. As it turns out, these particles, performing transverse oscillations, follow the designed tunnels of low irradiance. On the contrary, for particles with positive polarizability these areas are inaccessible.
\end{abstract}

\maketitle

\section{Introduction}\label{intr}

As it is well known, if polarization effects do not play a significant role, a laser beam near the propagation axis can be effectively described by the simplified scalar Helmholtz equation. The approximation is accomplished by first substituting into the wave equation 
\begin{equation}
\Big(\mathcal{4}_\perp +\partial^2_z-\frac{1}{c^2}\,\partial^2_t \Big)\Psi(\bm{r},z,t)=0,
\label{dal}
\end{equation}
where $\mathcal{4}_\perp$ denotes the transverse, two-dimensional Laplace operator, the solution in the form 
\begin{equation}\label{epi}
\Psi(\bm{r},z,t)=e^{ik(z-c t)}\psi(\bm{r},z),
\end{equation}
where $\psi(\bm{r},z)$ is assumed to be slowly varying function of the coordinate $z$. The electric field is then related to $\Psi(\bm{r},z,t)$ in the standard way:
\begin{equation}\label{epie}
\bm{E}(\bm{r},z,t)=\bm{E}_0 \Psi(\bm{r},z,t),
\end{equation}
(upon satisfying the condition $\bm{E}_0\cdot \bm{\nabla}\Psi(\bm{r},z,t)\approx 0$) with $\bm{E}_0$ representing a constant vector. Bold symbol $\bm{r}$ above (and later $\bm{\xi}$), denotes the two-dimensional vector lying in the plane perpendicular to the direction of propagation, e.g. $\bm{r}=[x,y]$.

Upon neglecting the second order derivative with respect to $z$ due to
\begin{equation}
|\partial^2_z\psi|\ll |k\partial_z\psi|,
\label{parap}
\end{equation}  
where $\partial_z$ denotes $\partial/\partial_z$, one gets the so called {\em paraxial equation} for the scalar envelope $\psi(\bm{r},z)$~\cite{sie}:
\begin{equation}\label{paraxial}
\left(\mathcal{4}_\perp +2ik\partial_z\right) \psi({\bm{r}},z)=0.
\end{equation}
 This approximation was worked out in detail by Lax and collaborators in \cite{lax}.

The fundamental solution to this equation called the Gaussian beam (GB), has been known since 1966 (see \cite{kl}), and extensively investigated within, but also beyond the validity of the paraxial approximation (see for instance \cite{sie,davis,nemo,mw,sesh,gustavo,er,selina,saleh}). 

The principal feature of GB, contrary to the unphysical and idealized infinite plane waves, constitutes the inhomogeneity in the distribution of the wave intensity, and especially the presence of a narrowing, termed the beam waist, where the concentration of energy is maximal. This property enabled the trapping of neutral polarizable particles and the design of the so-called optical tweezers, i.e. gradient force traps \cite{ash,chu,miller,pad}.

The structure of GB is further enriched if it is endowed with the nonzero orbital angular momentum (OAM). In this case, the beam has a vortical nature: on its axis the irradiance drops to zero, and upon encircling it the phase changes by $2\pi n$, where $n\hbar$ denotes the value of the OAM. Such beams can be said to be ``hollow'' along the propagation axis. The surfaces of the constant phase are then of helical character.

The original optical tweezers operated due to gradient forces pulling particles of positive polarizability into areas of high wave intensity. In the similar way, the irradiance ``holes'' can serve as traps or guide lines for objects of negative polarizability such as atoms in the blue-detuned beams \cite{yina,yinb,xxu}. The identical effect is owed to the ponderomotive force acting on charged particles, such as electrons, originating from the inhomogeneous circularly polarized wave \cite{ibb1,barredo}.

As a recent experimental example of the use of so-called Dark Focus Tweezers one can recall the work \cite{almeida} which demonstrates trapping of nanometer-sized silicon balls placed in an external medium to achieve a relative refractive index less than one. A hollow beam called a ``bottle beam'' \cite{melo}  was obtained in this study by superposition of Gaussian and Laguerre-Gaussian beams. Numerous references to other theoretical and experimental findings can be found in the works cited above.

Dark Focus Tweezers or hollow beams have an advantage that particles are trapped or guided in the low radiation domain, and are therefore less likely to be damaged. This is particularly significant for biological objects.
 
In this paper, we would like to focus on certain ``hollow'' Gaussian beams, which represent the solutions to the paraxial equation, and which can be designed to suit specific purposes, such as guiding several particles or atoms in a special way. Hollow beams have been for years of interest to researchers due to their possible applications in particle trapping, but also in atomic physics or optical communication (see for instance \cite{bal,xu,gao,cai,kha}). As said above, however, typically this term refers to the situation in which light is concentrated outward on a cylindrical or annular structure with a hollow space in the center aligned along the propagation axis (certain non-cylindrical hollow beams of elliptical or rectangular cross-section were introduced in \cite{cai}). To this category belong well-known Bessel-Gaussian or Laguerre-Gaussian beams but also other ones (see for instance \cite{saleh,gori,april1,mendoza,sie,lg,april2,nas,trsh,kuga,yin,yan,sun}). Contrary to this, the beams dealt with in the present work are generated as superpositions of two or more coaxial Gaussian modes with specific OAM values, that lead to the development of a {\em multi-hole} structure. Naturally, a superposition of cylindrical waves with different angular momenta is no longer a cylindrical beam in the sense that the distribution of the irradiance does not exhibit axial symmetry. Depending on the choice of the constituent modes, this multi-tube (i.e. multi-hole) structure, can be designed as needed. This issue will be discussed in detail in the next section.

This type of beams has been known in the literature for about 30 years, although it seems that not in the context of guiding particles. To mention just a few results one can first recall the work \cite{indeb}, which, after a rather general introduction containing theoretical foundations of the so called arrays of vortices, provides numerical results concerning vortex-vortex or vortex-antivortex interactions. In turn the experimental work \cite{bast} demonstrates the generation of a two-hole beam thanks to the diffraction on the computer-synthesized grating and examins the stability of nearly located vortices. In \cite{rozas} the authors focus on the numerical analysis of the motion and interaction of vortices inserted in a Gaussian beam, in a linear and nonlinear media. 

This kind of structured light is expected to find potential application in communication. For this reason, the stability of hollow beams subjected to random phase distorsions has also been investigated \cite {koval}.

In Section \ref{gpar} the possible use of this structure to transport particles in a certain way will be addressed. For example, a multi-hole beam can simultaneously guide several particles, each in its own potential tube. This paper is concerned with a theoretical, qualitative rather than quantitative description of this phenomenon.

Among all the research areas mentioned above, the manipulation of particles still remains a key problem because of its wide potential usage in physics, chemistry, biology or medicine (see for instance~\cite{pad,ste,fazal,woe,bowpa,grier1,brad}) and constitutes one of the major applications of the structured light. In particular, apart from $3D$ traps, guiding particles by light along pre-designed trajectories, as in \cite{liang,trknot,dxu}, continues to be an exciting topic, and both purely theoretical and experimental research in this area seems to be of importance.

\section{Description of multi-hole Gaussian beams}\label{descr}

Before proceeding, it is convenient to introduce dimensionless coordinates according to the formulas
\begin{equation}\label{newpara}
\xi_x=\frac{x}{w_0},\qquad \xi_y=\frac{y}{w_0},\qquad \zeta=\frac{z}{z_R},
\end{equation}
where $w_0$ is the beam waist and $z_R=\frac{kw_0^2}{2}$ denotes the Rayleigh length, i.e., the distance at which the area of the transverse section of the beam increases twice. A different designation is used for the transverse components ($\xi_x,\xi_y$) and for the longitudinal component ($\zeta$), as they play a somewhat different role in the subsequent expressions. With this notation the paraxial equation (\ref{paraxial}) takes the following form
\begin{equation}\label{parrxz}
\left(\mathcal{4}_{\xi\perp} +4i\partial_\zeta\right) \psi({\bm{\xi}},\zeta)=0
\end{equation}
where $\bm{\xi}=[\xi_x,\xi_y]$. The solution can be looked for in the following form
\begin{equation}\label{zmpsi}
\psi(\bm{\xi},\zeta)=\frac{1}{1+i\zeta}e^{-\frac{\bm{\xi}^2}{1+i\zeta}}\,\tilde{\psi}({\bm{\xi}},\zeta).
\end{equation}

\begin{figure}[h!]
\centering
\includegraphics[width=0.45\textwidth,angle=0]{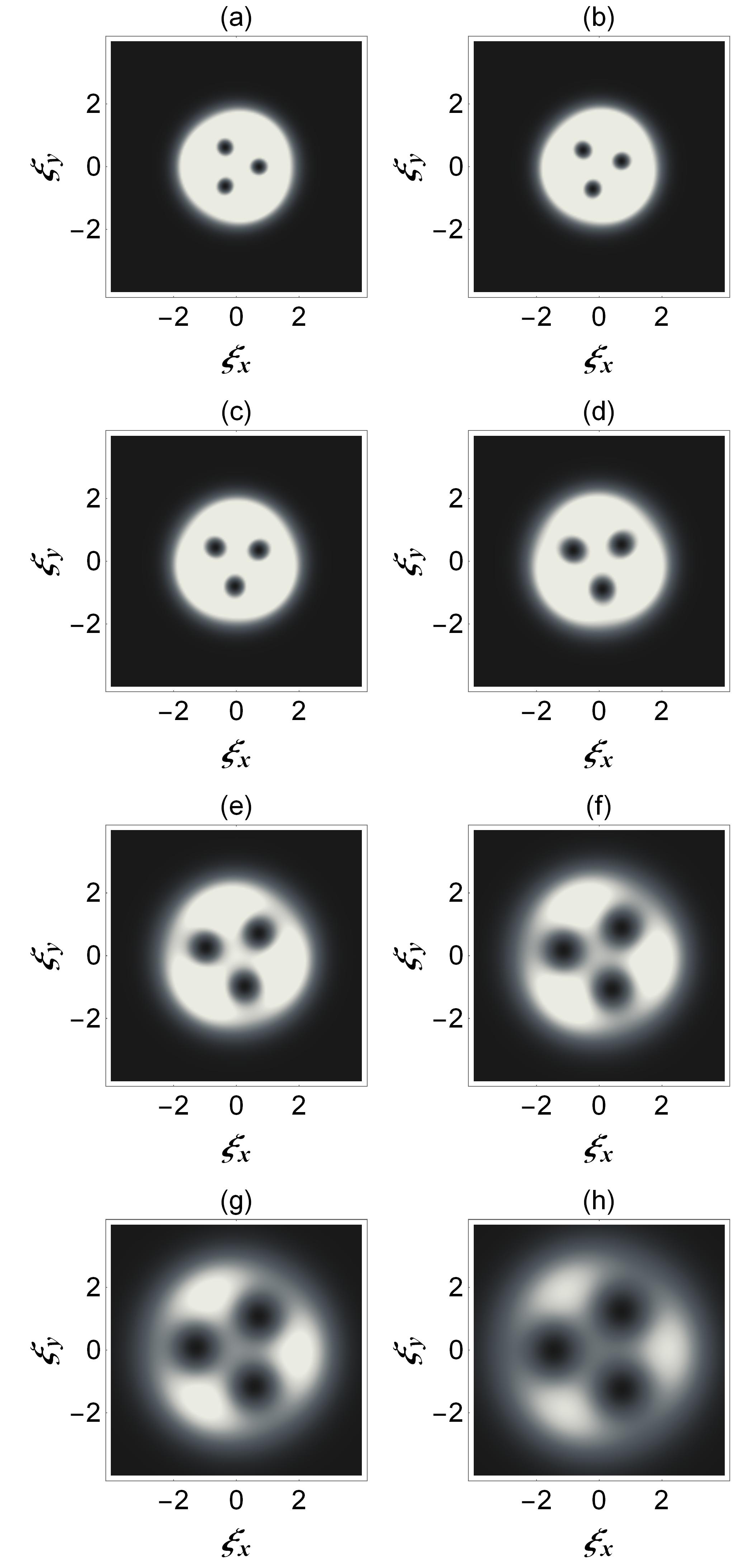}
\caption{The irradiance, in the perpendicular plane, of the beam created as the plain superposition of (\ref{zmpsin}) and (\ref{zmpsinn}) for $n=3$ according to (\ref{poli}). The following values of $\zeta$ are used in subsequent plots: a) $0$, b) $0.25$, c) $0.5$, d) $0.75$, e) $1$, f) $1.25$, g) $1.5$, h) $1.75$. The parameter $\beta= 1.4$.  Bright areas represent high irradiance and dark ones low irradiance.}
\label{xyhole3}
\end{figure}

After the substitution of (\ref{zmpsi}) into (\ref{parrxz}) one can easily derive the differential equation for the unknown function $\tilde{\psi}({\bm{\xi}},\zeta)$:
\begin{equation}\label{xzxa}
\xi\partial_\xi\tilde{\psi}(\xi,\zeta)=i(1+i\zeta)\partial_\zeta\tilde{\psi}(\xi,\zeta),
\end{equation}
which is satisfied by any function of one combined argument $\frac{\xi}{1+i\zeta}$, i.e. 
\begin{equation}\label{zmpt}
\tilde{\psi}=\tilde{\psi}\left(\frac{\xi}{1+i\zeta}\right).
\end{equation}
In the above formulas (\ref{xzxa}) and (\ref{zmpt}) the complex coordinate $\xi=\xi_x+i\xi_y$ is introduced and should be distinguished from $|\bm{\xi}|=\sqrt{\xi_x^2+\xi_y^2}=|\xi|$. Result (\ref{zmpt}) is well known~\cite{abra}. In particular the choice of $\tilde{\psi}(s)=\mathrm{const}$, leads to the fundamental GB, and $\tilde{\psi}(s)=\mathrm{const}\cdot s^n$ corresponds to the GB of vorticity $n$.

\begin{figure}[h!]
\centering
\includegraphics[width=0.45\textwidth,angle=0]{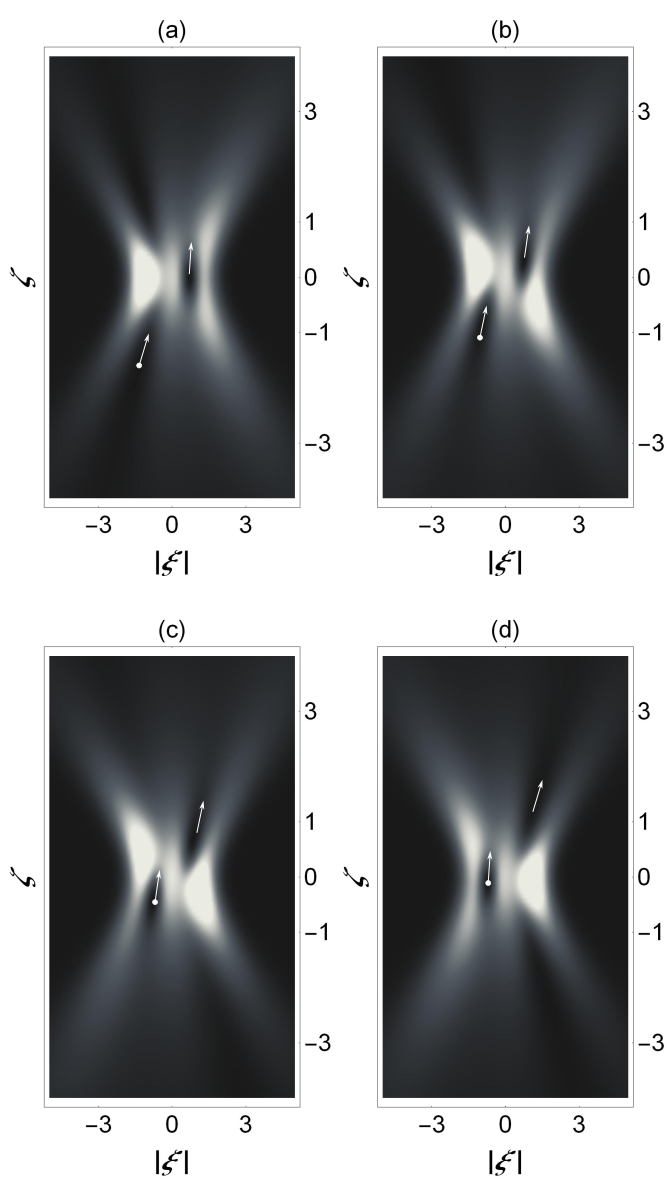}
\vspace{-4.5ex}
\caption{Same as Fig. \ref{xyhole3} but in the axial plane. Subsequent graphs present the view on the same beam from different angles between the $x$ axis and the line of sight : a) $0$, b) $0.1\pi$, c) $0.2\pi$, d) $0.3\pi$. White arrows of two kinds mark two twisting tubes of low irradiance.}
\label{xzhole3}
\end{figure}

Let us now concentrate on the zeros of the envelope $\psi$. From the formula (\ref{zmpsi}) it is obvious that it does not have any other zeros than those of the function $\tilde{\psi}$. Let $p=p_r+ip_i=p_0e^{i\phi_0}$ be any of them. As $\zeta$ increases, i.e. when moving upward along the beam, the radial distance of this zero from the beam axis grows, which is consistent with the diffraction of the beam itself, but also its position gets twisted around it by an angle asymptotically tending to $\pi/2$, as indicated in the formulas: 
\begin{equation}\label{strlin}
\left\{\begin{array}{l}|\bm{\xi}|=\displaystyle\frac{p_0}{\beta}\sqrt{1+\zeta^2},\\ \phi(\zeta) =\phi_0+\arctan\zeta,\end{array}\right. 
\end{equation}
where $\beta$ is a certain constant playing the role of the scaling factor defined below in (\ref{poli}).
Since all possible zeros follow synchronized, identical paths, they never merge, their number remains constant along the beam, and each develops its own nodal line as the value of $\zeta$ increases (naturally in the opposite direction, i.e. for $\zeta<0$ as well). This effect was observed in \cite{indeb}.

The simplest function that can be picked is a polynomial, which corresponds to the interference of a couple of coaxial Gaussian beams with differing OAM values. Of course, $\tilde{\psi}$ is at our disposal and any analytic function could play this role, although a non-polynomial function would require a superposition of infinitely many modes (from the practical point of view, however, low-intensity high-order beams might be ignored). For the purposes of this paper, the function $\tilde{\psi}$ is chosen in the form of a particularly simple polynomial:
\begin{equation}\label{poli}
\tilde{\psi}(s)=(\beta s)^n-1.
\end{equation}
The scaling role of the parameter $\beta$ has already been mentioned and is clear in (\ref{poli}). This form indicates that the interference of exactly two Gaussian beams is dealt with: that of order $0$:
\begin{equation}\label{zmpsin}
\psi_0(\bm{\xi},\zeta)=\frac{1}{1+i\zeta}e^{-\frac{\bm{\xi}^2}{1+i\zeta}},
\end{equation}
and the other of order $n$:
\begin{equation}\label{zmpsinn}
\psi_n(\bm{\xi},\zeta)=\frac{\beta^n\xi^n}{(1+i\zeta)^{n+1}}e^{-\frac{\bm{\xi}^2}{1+i\zeta}}.
\end{equation}
In our analysis the overall normalization constants are omitted, as only relative beam intensities are essential (in a sense represented by the value of the parameter $\beta$).

Beam (\ref{poli}) bears the vortex topological charge $n$ and when encircling the axis $\zeta$ the value of the phase increases by $2\pi n$ which means that it assumes $n$ times the same values (if reduced to the interval $[0, 2\pi[$). Thus, on a circle (for $\zeta=\mathrm{const}$) of radius $\sqrt{1+\zeta^2}/\beta$ (in units of $w_0$) a completely destructive interference with $\tilde{\psi}_0$ occurs exactly $n$ times. The $n$th degree vortex ``spreads'' into $n$ individual vortices (in the case of Bessel beams the same phenomenon was demostrated in \cite{berrya}), uniformly distributed, as is obvious from the distribution of the $n$th complex roots of the unity. Consequently, there appear $n$ ``holes'' in the wave intensity in the perpendicular plane, as shown in Fig. \ref{xyhole3} for $n=3$. In the subsequent diagrams (a)-(h) performed for increasing values of $\zeta$, the diffraction of the wave and the twisting of the whole pattern can be seen, in agreement with (\ref{strlin}).

In Fig. \ref{xzhole3} the irradiance of the same beam in four axial planes is drawn. The first plane is simply $\xi_x\zeta$ plane, and the subsequent ones are rotated around $\zeta$-axis by the successive multiples of $\pi/10$. The formation of the zero-intensity tubes is marked with arrows. 

\begin{figure}[h!]
\begin{center}
\includegraphics[width=0.45\textwidth,angle=0]{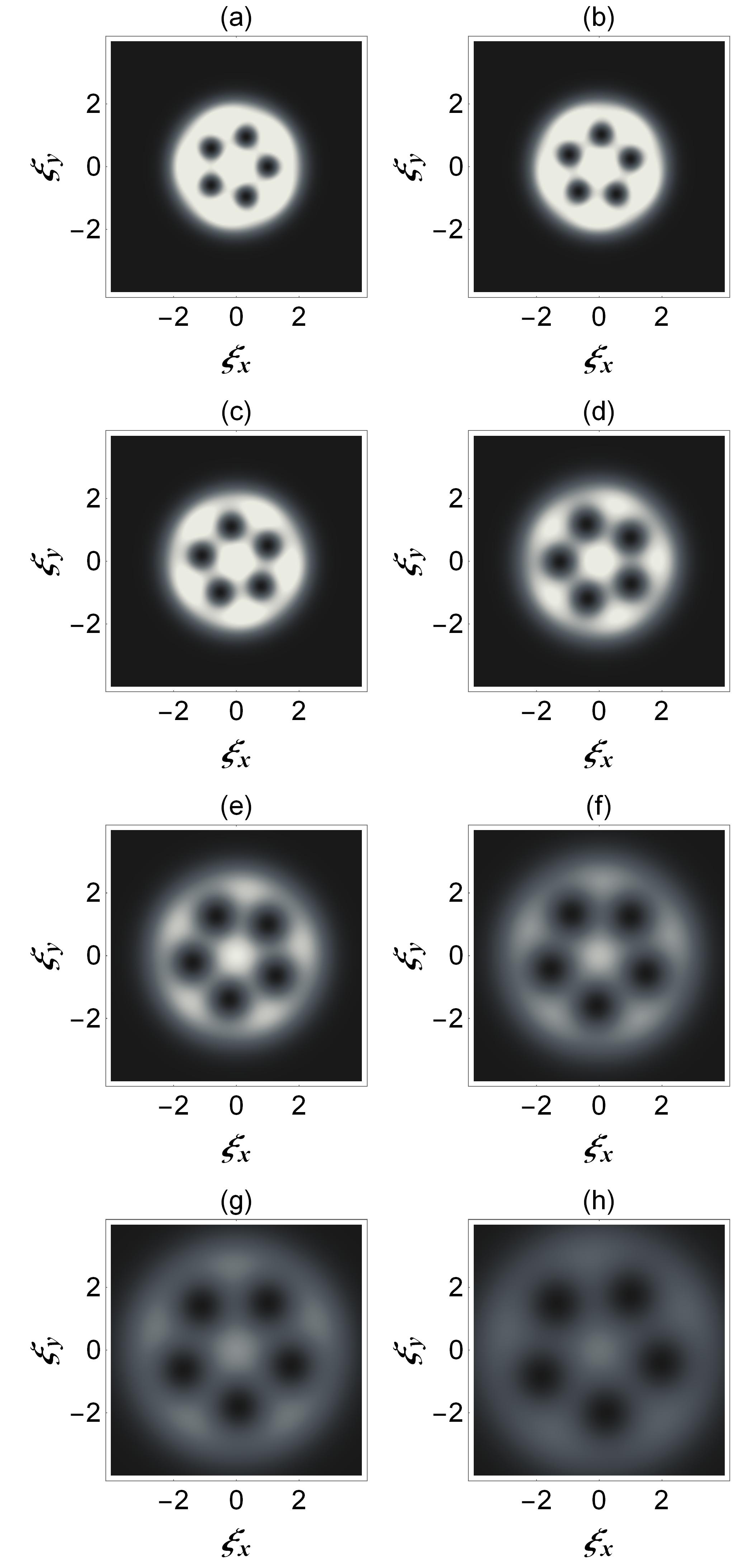}
\end{center}
\vspace{-4.5ex}
\caption{Same as Fig. \ref{xyhole3}, but for $n=5$ and $\beta=1$.}
\label{xyhole5}
\end{figure}

\begin{figure}[h!]
\begin{center}
\includegraphics[width=0.45\textwidth,angle=0]{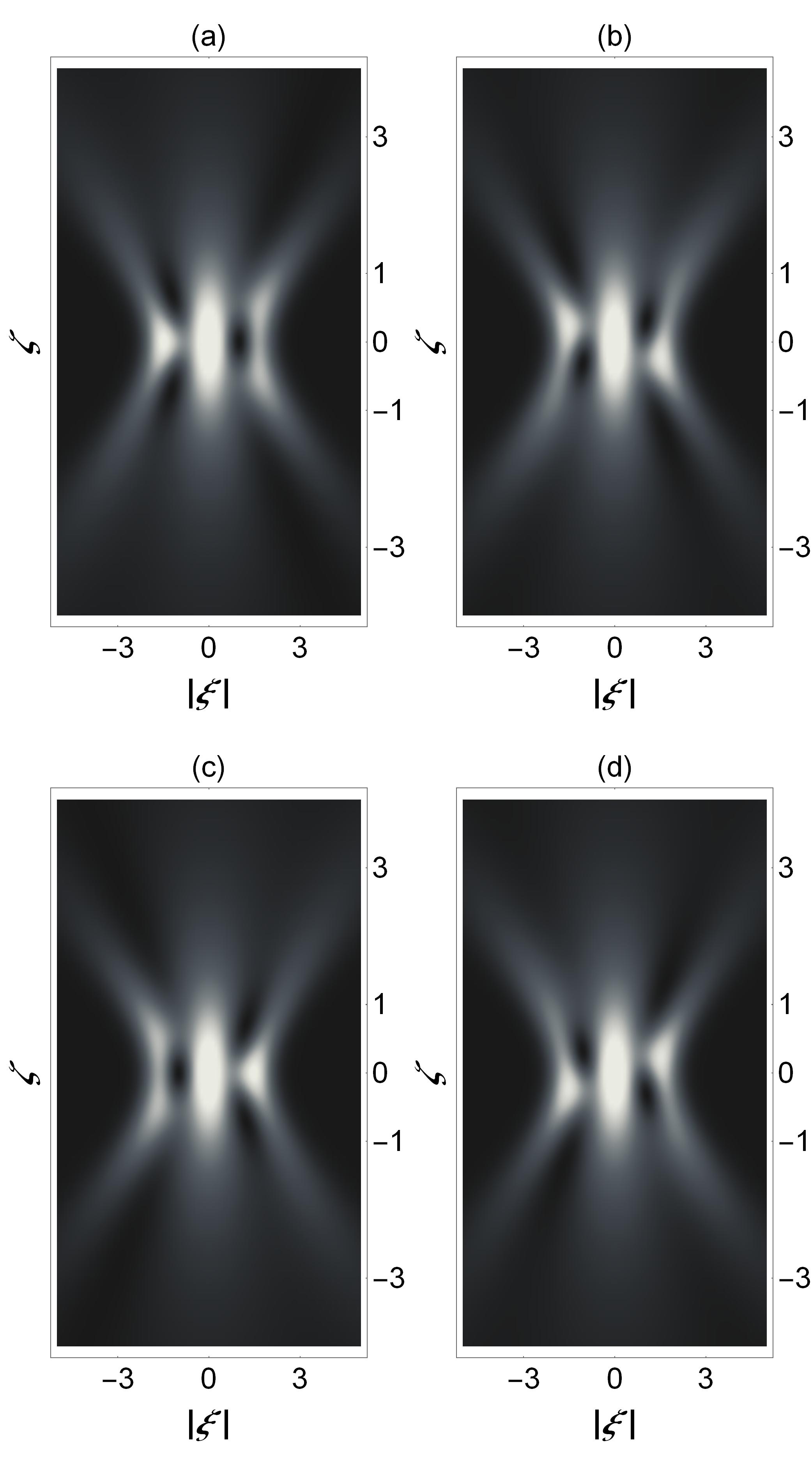}
\end{center}
\vspace{-4.5ex}
\caption{Same as Fig. \ref{xzhole3} but for $n=5$ and $\beta=1$.}
\label{xzhole5}
\end{figure}

Figures~\ref{xyhole5} and~\ref{xzhole5} demonstrate the same effects for $n=5$. The existence of five distinct roots of unity yields five ``tubes'' of vanishing irradiance that can eventually be exploited. A slightly reduced value for the parameter $\beta$ has been chosen in this case in order to avoid merging areas of low energy density.

Naturally, in the role of $\tilde{\psi}$ other polynomials, that would not have zeros distributed in such a regular fashion, come into play as well.

In order to visualize the splitting of one vortex of higher topological charge into several single ones it is convenient to analyze the phase of the beam in the perpendicular planes. Fully destructive interference requires (apart from the equality of the wave amplitudes) the phases of the two interfering waves at a given point to differ by an odd multiple of $\pi$. It is known that at such places, the overall phase $\phi$ of the combined wave becomes indeterminate. 

\begin{figure}[h!]
\begin{center}
\includegraphics[width=0.45\textwidth,angle=0]{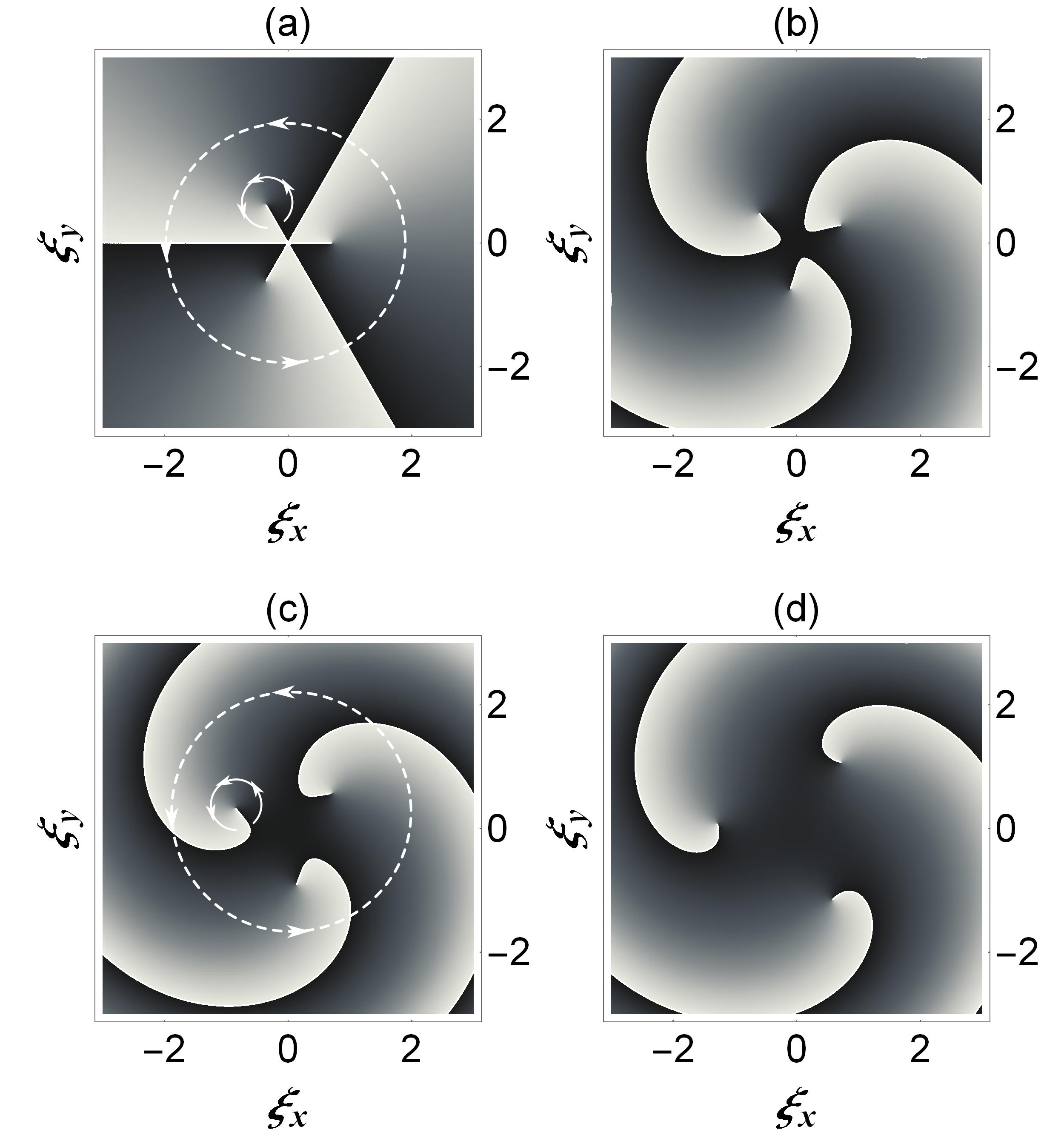}
\end{center}
\vspace{-4.5ex}
\caption{The phases of the wave-function of Fig. \ref{xyhole3} depicted in four planes: a) $\zeta = 0$, b) $0.4$, c) $0.8$, d) $1.5$. The value of the phase, modulo $2\pi$, is represented continuously by means of the grayscale from $-\pi$ (black color) to $\pi$ (white color). The rotation of the entire picture with increasing $\zeta$ is owed to the factor $1+i\zeta$ in (\ref{zmpsin}) and(\ref{zmpsinn}) and the additional factor $e^{ikz}=e^{i\zeta}$ coming from \ref{epi}. The small white circle represents the curve ${\cal C}$ circulating around one of the individual vortices produced by the breakdown of the vortex with topological charge $n=3$. The large white circle drawn with the dashed line encircles all the resultant vortices.}
\label{xyphase3}
\end{figure}

The change of phase along a certain closed curve ${\cal C}$ is defined by the formula 
\begin{equation}\label{kfaza}
\Delta_{\cal C}\phi=\oint\limits_{\cal C}\bm{\nabla}\phi\, d\bm{l}=-\frac{i}{2}\oint\limits_{\cal C}\frac{\Psi^*\stackrel{\leftrightarrow}{\bm{\nabla}}\Psi}{\Psi^*\Psi}\, d\bm{l}
\end{equation}
where $*$ denotes the complex conjugation. In the case dealt with here the integration contour can be deformed to the flat one lying in the plane $\zeta=\mathrm{const}$, such as white circles drawn in Fig.~\ref{xyphase3}.  Then the nabla operator reduces to two dimensions (vector $d\bm{l}$ does not have the $\zeta$ component) and $\Delta_{\cal C}\phi$ can be represented in the form of the complex contour integral with respect to $d\xi=d\xi_x+id\xi_y$: 

\begin{eqnarray}
\Delta_{\cal C}\phi&=&-\frac{i}{2}\oint\limits_C\left(\frac{\bm{\nabla}\psi}{\psi}-\frac{\bm{\nabla}\psi^*}{\psi^*}\right)\, d\bm{l}\label{kofaza}\\
&=&-\frac{i}{2}\oint\limits_C\Big[\frac{1}{\psi}(d\xi_x\,\partial_{\xi_x}\psi+d\xi_y\,\partial_{\xi_y}\psi)\nonumber\\
&&\;\;\;\;-\frac{1}{\psi^*}(d\xi_x\,\partial_{\xi_x}\psi^*+d\xi_y\,\partial_{\xi_y}\psi^*)\Big]\nonumber
\end{eqnarray}

\begin{figure}[h!]
\begin{center}
\includegraphics[width=0.45\textwidth,angle=0]{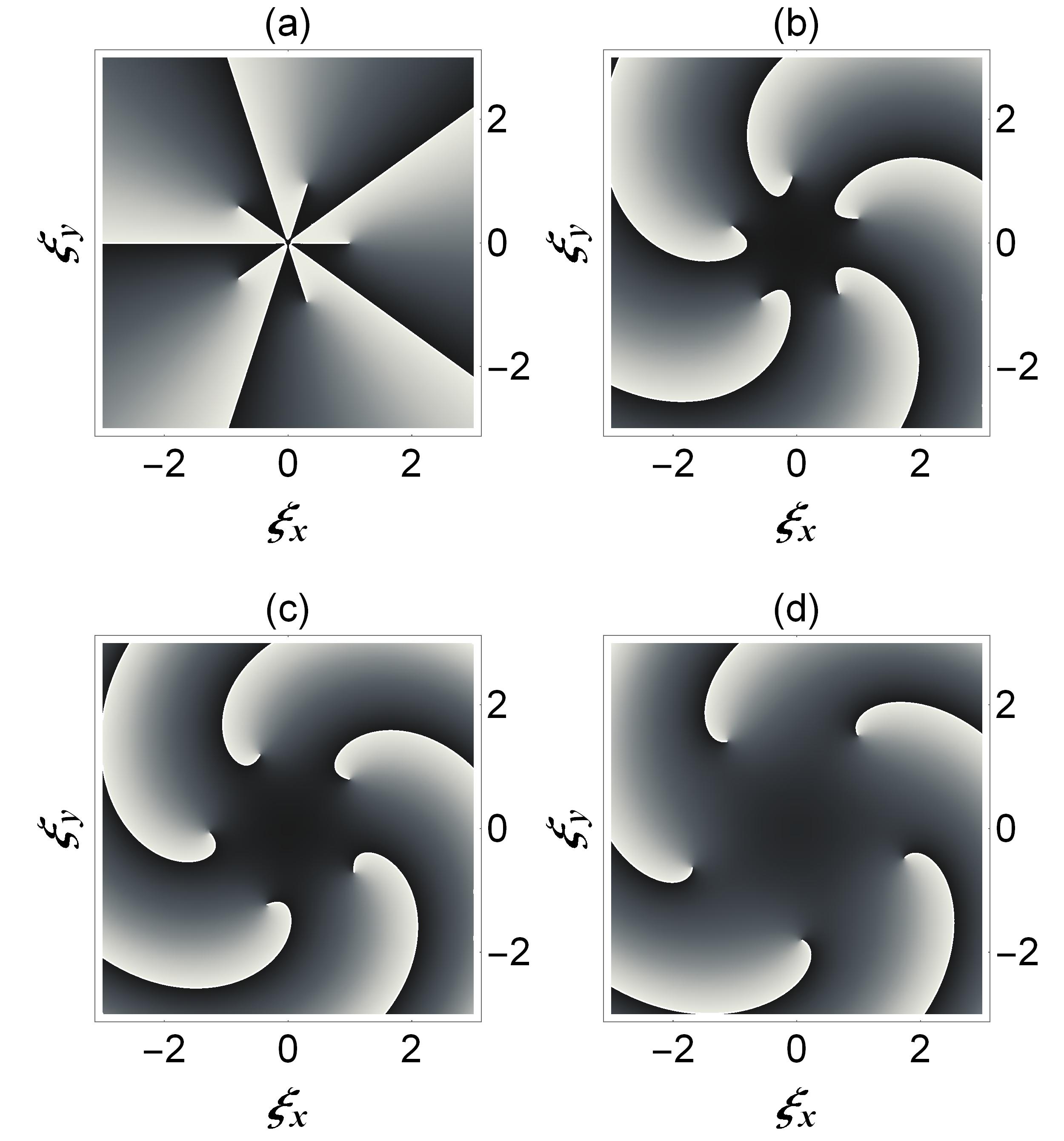}
\end{center}
\vspace{-4.5ex}
\caption{Same as Fig. \ref{xyphase3} but for $n=5$ and $\beta=1$.}
\label{xyphase5}
\end{figure}

The function $\psi$ formally depends on two complex variables: $\xi$ and $\xi^*$, so 
\begin{equation}
\partial_{\xi_x}\psi=(\partial_\xi+\partial_{\xi^*})\psi,\qquad \partial_{\xi_y}\psi=i(\partial_\xi-\partial_{\xi^*})\psi,
\end{equation}
and identically for $\psi^*$. Expression (\ref{kofaza}) may be then rewritten as
\begin{eqnarray}
\Delta_{\cal C}\phi&=&-\frac{i}{2}\oint\limits_C\Big[\frac{\partial_\xi\psi}{\psi}\, d\xi+\frac{\partial_{\xi^*}\psi}{\psi}\, d\xi^*\nonumber\\
&&\;\;\;\;-\frac{\partial_\xi\psi^*}{\psi^*}\, d\xi-\frac{\partial_{\xi^*}\psi^*}{\psi^*}\, d\xi^*\Big]\label{roxi}
\end{eqnarray}
The value of this integral can be obtained either by direct substitution or via the Cauchy argument principle. Formally $\psi$ is not holomorphic in any domain as it depends on $\xi^*$. However, for functions $\psi$ of the form $f(\xi)e^{-a\xi\xi^*}$, the terms in which the exponential is subject to differentiation do not contribute, since
\begin{equation}\label{znikx}
\oint\limits_C(\xi^*d\xi+\xi d\xi^*)=0.
\end{equation}
In all other terms (i.e. those in which the exponential is not differentiated) the exponentials in numerator and denominator cancel out and the trace of $\xi^*$ disappears from the expression. Therefore, form the practical point of view the function $\psi$ may be treated as holomorphic and this is how it is handled below. In general, this argumentation does not necessarily apply for beams for which the function $f$ does also depend on $\xi^*$, such as Hermite-Gaussian beam~\cite{kl,sie} or that of~\cite{trgen}.

Consequently, keeping in mind (\ref{znikx}), we can write
\begin{equation}
\Delta_{\cal C}\phi=-\frac{i}{2}\left[\oint\limits_C\frac{\partial_\xi\psi}{\psi}\,d\xi-\left(\oint\limits_C\frac{\partial_\xi\psi}{\psi}\,d\xi\right)^*\right],\label{ghte}
\end{equation}
From the Cauchy argument principle it stems that
\begin{equation}\label{zpo}
\oint\limits_C\frac{\psi'}{\psi}\, d\xi= 2\pi i(Z-P)
\end{equation}
where $Z$ denotes the number of zeros and $P$ the number of poles in the area encompassed by the curve ${\cal C}$. However $\psi$ has no poles, and the only zeros come from the polynomial (\ref{poli}), and hence we come to
\begin{equation}\label{zrepol}
\oint\limits_C\bm{\nabla}\phi\, d\bm{l}=2\pi Z.
\end{equation}
Since this polynomial has $n$ single zeros, for the integral over small white circles of Fig. \ref{xyphase3} one always gets the value of $2\pi$, and for the large one $2\pi n$ (here $n=3$), which means that the total vorticity is unchanged, but the vortex merely gets splitted into $n$ single vortices. Naturally, the same can be observed in Fig.~\ref{xyphase5}.

\section{Guiding of particles}\label{gpar}

As is well known, inhomogeneities in the intensity of the wave, and hence in the electric field, can be exploited to trap and guide neutral particles, for instance atoms or dielectric balls, which undergo polarization in external fields. This has become the basis for the operation of the so-called optical tweezers \cite{ash,chu,miller,pad}, as mentioned in Introduction.

Let us denote with $\alpha$ the particle polarizability, which in general can depend on the driving frequency, and with $d$ the induced dipole moment. Then
\begin{equation}\label{de}
{\bm d}=\alpha {\bm E}.
\end{equation}
When speaking of atoms, from the theory of the AC Stark effect it is known that for a red-detuned beam the atomic polarizability $\alpha$ is positive and for blue-detuned one it becomes negative, which affects the motion of atoms in a fundamental way. Similar conclusions can be drawn from a purely classical model of the atom~\cite{kra,odtna}. The analogous efect in the case of dielectric nano-spheres is owed to the value of the refractive index higher or lower than that of the surrounding media.

The Newton equation of motion of an atom (or other particle) in these conditions takes the form
\begin{equation}
m\ddot{\bm r}= ({\bm d}\cdot {\bm \nabla}){\bm E}=\frac{1}{2}\,\alpha {\bm \nabla}(E^2)+\bm{F}_{\mathrm{scat}},
\label{rr1}
\end{equation}
the right hand side of the equation being treated as averaged over the fast oscillations of the field in (\ref{epi}). The first term represents the conservative gradient force and $\bm{F}_{\mathrm{scat}}$ stands for the scattering force to be dealt with below in some detail. 

The irradiance in this kind of beams varies rapidly in directions orthogonal to the beam axis and relatively slowly alongside it, so the perpendicular components of $\bm{F}_{\mathrm{scat}}$ can be ignored compared to the gradient force and only parallel one needs to be accounted for, which is confirmed below. In order to estimate the magnitude of this force we will focus on the example of dielectric nano-spheres. The influence of this force on atoms can be reduced by using far-detuned beam, although it should be kept in mind that the gradient force is also weakened in this way.

As it is known \cite{kerker,harada} for dielectric spheres one has
\begin{equation}\label{cofor}
\frac{F_{\mathrm{scat}\,z}}{F_{\mathrm{grad}\,z}}=\frac{n_sk^4\alpha}{3\pi n_m^3\varepsilon_0}\,\frac{I}{\partial_z I},
\end{equation}
where $n_s$ is the refractive index of the sphere, $n_m$ -- the refractive index of the surrounding media, $\varepsilon_0$ -- the vacuum permittivity and $I$ stands for the irradiance. In dimensionless coordinates $\xi_x,\xi_y,\zeta$ defined in (\ref{newpara}) this ratio can be given the form
\begin{equation}\label{coforb}
\frac{F_{\mathrm{scat}\,\zeta}}{F_{\mathrm{grad}\,\zeta}}=\frac{n_s}{n_m^3}\, \zeta_R\kappa\,\frac{I}{\partial_\zeta I}
\end{equation}
with $\zeta_R=k z_R$ and $\kappa=\frac{k^3\alpha}{3\pi\varepsilon_0}$. Let us estimate these quantities for Rayleigh particles placed in a tightly focused light beam with the waist radius $w_0=0.05\,\mathrm{mm}$ and wavelength $\lambda=700\,\mathrm{nm}$ for which $\zeta_R\approx 10^5$. In order to find the magnitude of $\kappa$ one has to think of a concrete sample of particles. Take, for example, magnesium fluoride nano-balls ($n_s=1.38$) of diameter $D=10\,\mathrm{nm}$ placed in the glycerol ($n_m=1.47$). In these conditions one gets $\kappa\approx -1.1\cdot 10^{-5}$, where the well known formula for the polarizability \cite{harada} has been used:
\begin{equation}\label{alphare}
\alpha=\frac{1}{2}\,n_m^2\varepsilon_0 D^3\,\frac{n_s^2/n_m^2-1}{n_s^2/n_m^2+2}.
\end{equation}
Consequently the whole factor in (\ref{coforb}) equals
\begin{equation}\label{tofac}
\frac{n_s}{n_m^3}\, \zeta_R\kappa\approx -0.48,
\end{equation}

As regards the quotient $\frac{I}{\partial_\zeta I}$, it should be kept in mind that inside the ``tube'' of low irradiance, $I$ drops to zero (together with the scattering force). Naturally, the same refers to $\partial_\zeta I$ (since the vortex core constitutes the irradiation minimum) but the latter decline is weaker, as it is shown below. 

For the beams dealt with in this work, defined in (\ref{poli}), (\ref{zmpsin}) and (\ref{zmpsinn}) the needed ratio can be calculated to be
\begin{eqnarray}
\left|\frac{I}{\partial_\zeta I}\right|&=&\left|\frac{|\Psi|^2}{\partial_\zeta|\Psi|^2}\right|
=\Bigg|-\frac{2(n+1)\zeta}{1+\zeta^2}+\frac{4|\xi|^2\zeta}{(1+\zeta^2)^2}\nonumber\\
&&-\sum_{l=0}^{n-1}\left(\frac{\partial_\zeta\xi_l}{\xi-\xi_l}+\frac{\partial_\zeta\xi_l^*}{\xi^*-\xi_l^*}\right)\Bigg|^{-1},\label{rati}
\end{eqnarray}
where 
\begin{equation}\label{xkj}
\xi_l=\frac{1}{\beta}(1+i\zeta)e^{\frac{2\pi i l}{n}},\quad \mathrm{and}\quad \partial_z\xi_l=\frac{i}{\beta}\,e^{\frac{2\pi i l}{n}}.
\end{equation}
For particles guided in the $i$th tube $|\xi-\xi_i|\ll 1/\beta$ and the $i$th term in the sum in(\ref{rati}) dominates over all other ones. Therefore
\begin{equation}\label{stio}
\frac{F_{\mathrm{scat}\,\zeta}}{F_{\mathrm{grad}\,\zeta}}\approx \frac{n}{n_s^3}\, \zeta_R\kappa\,\left|\frac{\xi-\xi_i}{\partial_\zeta\xi_i}+\frac{\xi^*-\xi_i^*}{\partial_\zeta\xi_i^*}\right|\ll 1,
\end{equation}
and the scattering force should not have a relevant effect on the motion of guided particles. 

This estimate has been carried out for tightly focused beams. They can be broadened at the price of increasing the scattering force (it grows quadratically with $w_0$ provided the local energy density is maintained). However, this has no significant effect on the trajectories of guided particles (there is a quantitative effect, but not a qualitative one), as tested numerically. 

The scattering force has a little more significant effect for wider trajectories that depart from the core of vanishing irradiance but becomes truely essential for particles of positive polarizability which avoid areas of low irradiance. The motion of this type of particles is also shown in several figures below, therefore $F_{\mathrm{scat}\,\zeta}$ is taken into account in all numerical calculations.

\begin{figure}[h!]
\begin{center}
\includegraphics[width=0.45\textwidth,angle=0]{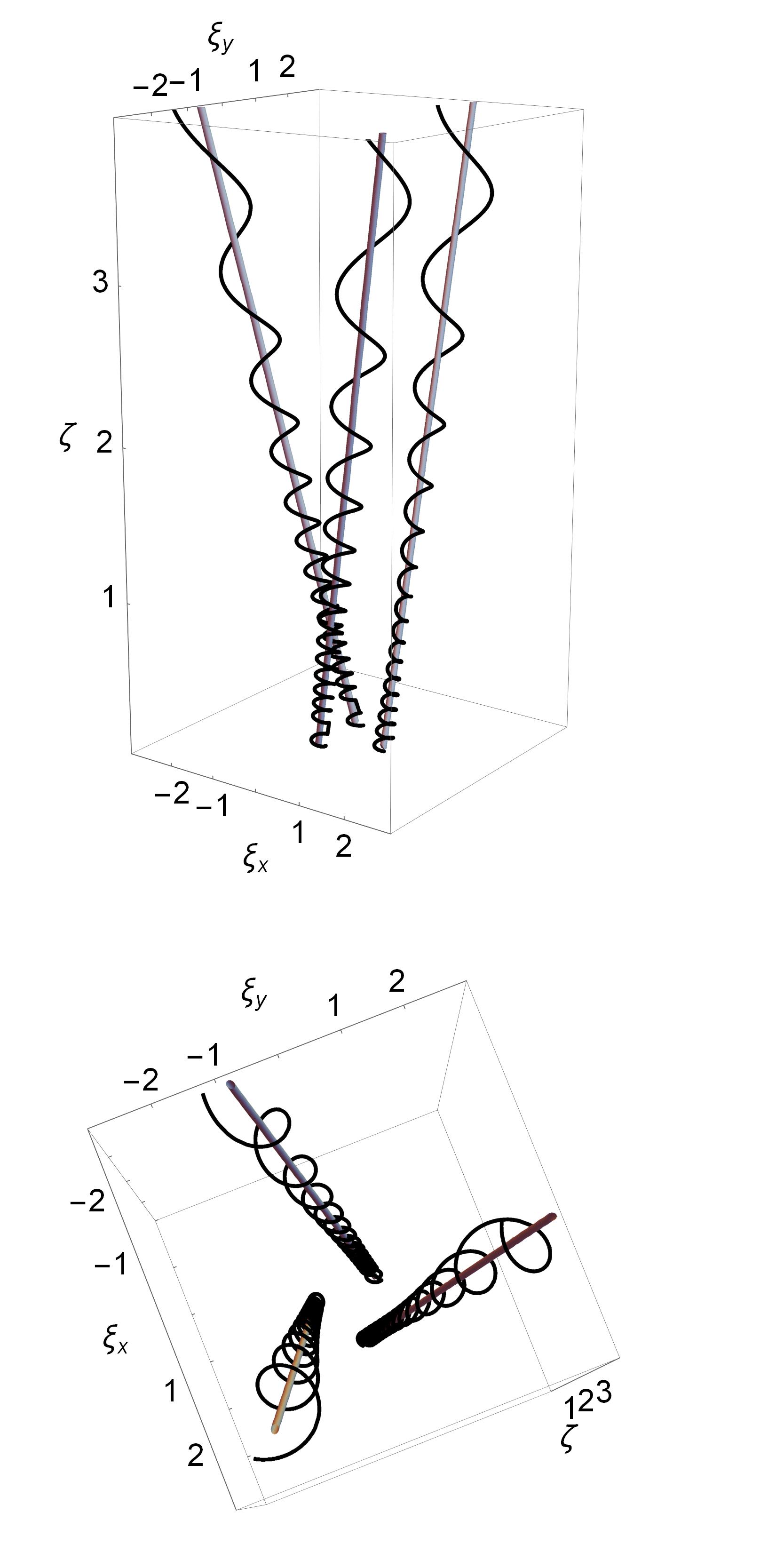}
\end{center}
\vspace{-4.5ex}
\caption{The trajectories of three particles of negative polarizability placed in the ``holes'' of the beam in question for $n=3$, viewed from the side and from above. Straight lines represent zero irradiance tubes, which undergo diffraction. The values of beam parameters are the same as in Fig. \ref{xyhole3}, and $\gamma=6\cdot 10^{-2}$ and $\tilde{\gamma}=2\cdot 10^{-5}$.}
\label{traj3}
\end{figure}

Consequently the smoothed (with respect to time) equations of motion can be given the following form:
\begin{subequations}\label{ewxi}
\begin{align}
&\ddot{\xi}_x=\gamma\,\partial_{\xi_x}|\psi(\bm{\xi},\zeta)|^2,\label{ewxix}\\
&\ddot{\xi}_y=\gamma\,\partial_{\xi_y}|\psi(\bm{\xi},\zeta)|^2,\label{ewxiy}\\
&\ddot{\bm{\zeta}}=\tilde{\gamma}\left(\partial_\zeta |\psi(\bm{\xi},\zeta)|^2+\frac{n_s}{n_m^3}\, \zeta_R\kappa|\psi(\bm{\xi},\zeta)|^2\right),\label{ewxiz}
\end{align}
\end{subequations}
where the coefficients $\gamma$ and $\tilde{\gamma}$ are expressed through the beam's parameters and particle mass as follows
\begin{subequations}\label{gammy}
\begin{align}
&\gamma=\frac{\alpha E_0^2}{4w_0^2\omega^2m},\label{gammya}\\
& \tilde{\gamma}=\frac{\alpha E_0^2}{4z_R^2\omega^2m}=\gamma\left(\frac{2}{kw_0}\right)^2=\frac{2}{\zeta_R}\, \gamma,\label{gammyb}
\end{align}
\end{subequations}
ande $\omega=kc$.
Depending on the sign of $\alpha$ the tubes described in the preceding section constitute either some kind of potential ``valleys'' ($\gamma<0$) or repulsive-potential ``hills'' ($\gamma>0$). Naturally, it is impossible to derive analytical solutions to the equations of motion in this kind of potentials, but trajectories of particles can be found numerically, for certain illustrative $\gamma$-parameter value. 

In Fig. \ref{traj3} it can be observed, from two different perspectives, that the trajectories of three negatively polarizable particles, having been inserted into the beam shown in Figs. \ref{xyhole3}, \ref{xzhole3} and \ref{xyphase3}, follow the irradiance ``holes''. The value of $\gamma=0.06$ was chosen for visualization purposes. 
Note the expansion of the particle's trajectory as it moves along the potential tube. It is related to the lowering of the potential barrier, which is associated with the beam's diffraction and with some acceleration of the particle. This potential barrier decreases proportionally to $1/w(z)^2$ or to $1/(1+\zeta^2)$. Over a distance of one Rayleigh length it then decreases twice. The potential valley becomes shallower, and the trajectory proportionally wider. From the practical point of view, such a distance seems  sufficient.

\begin{figure}[h!]
\begin{center}
\includegraphics[width=0.4\textwidth,angle=0]{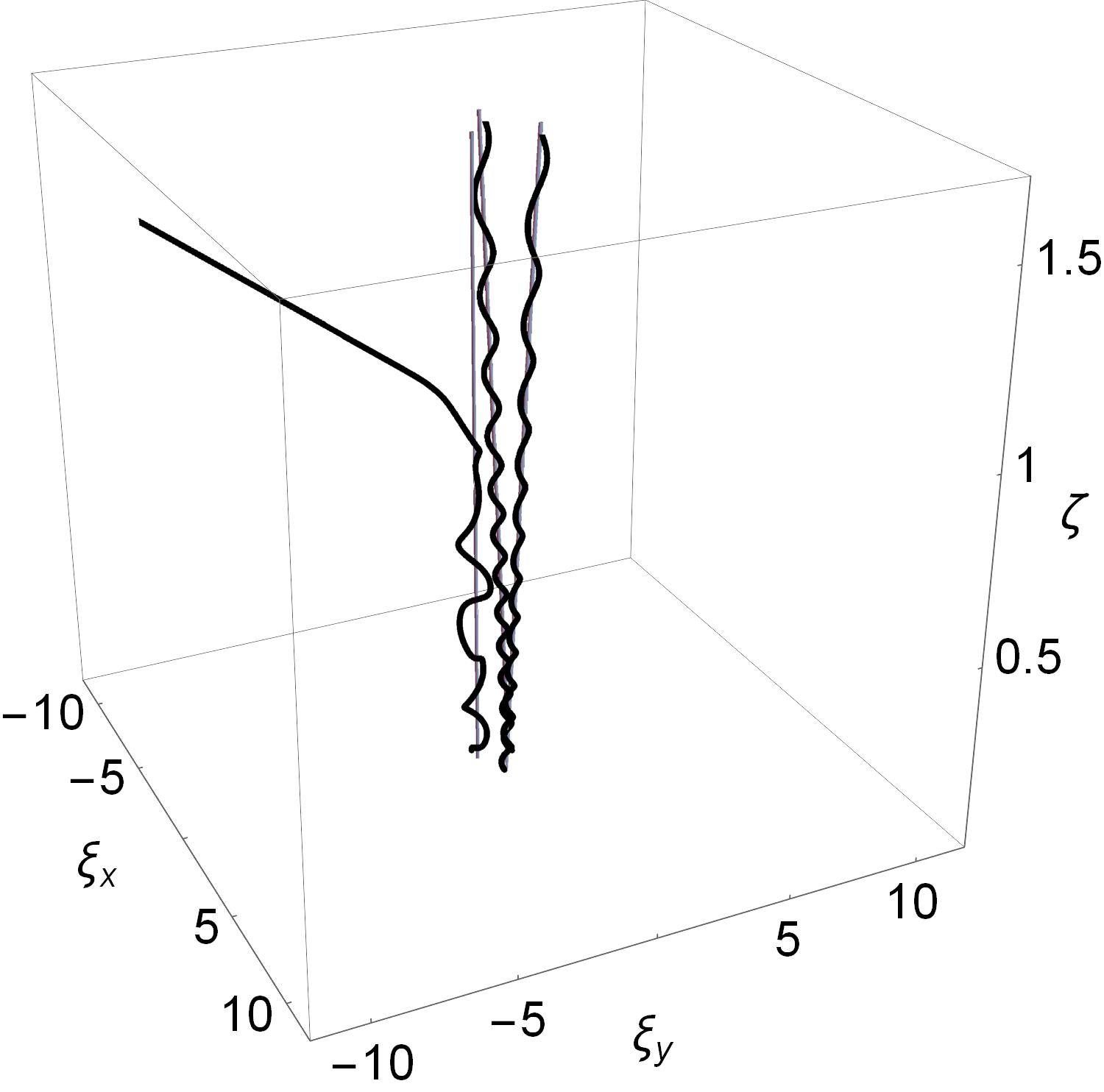}
\end{center}
\vspace{-4.5ex}
\caption{Escape of a particle from the potential tube. The values of parameters are the same as in Fig. \ref{traj3}.}
\label{esc}
\end{figure}

The lowering of the barrier height and width may be of importance for the eventual hopping of particles between different potential valleys. Especially in the quantum case, it affects the tunneling probability of guided elementary particles between various minima. Roughly speaking, if one neglects the possible acceleration of particles due to the gradient and scattering forces along the beam, the tunneling probability $p$ increases with $\zeta$ as $p^{(1+\zeta^2)^{-1/2}}$. 

Figure \ref{esc} shows the exemplary effect of the escape of one of the guided particles due to insufficient cooling and imprecise initial conditions which entail more chaotic motion. In \cite{trq} strong correlation between the chaoticness of classical trajectories of particles and the quantum tunneling probability in a trap constructed of a Bessel beam and constant magnetic field was established.

\begin{figure}[h]
\begin{center}
\includegraphics[width=0.45\textwidth,angle=0]{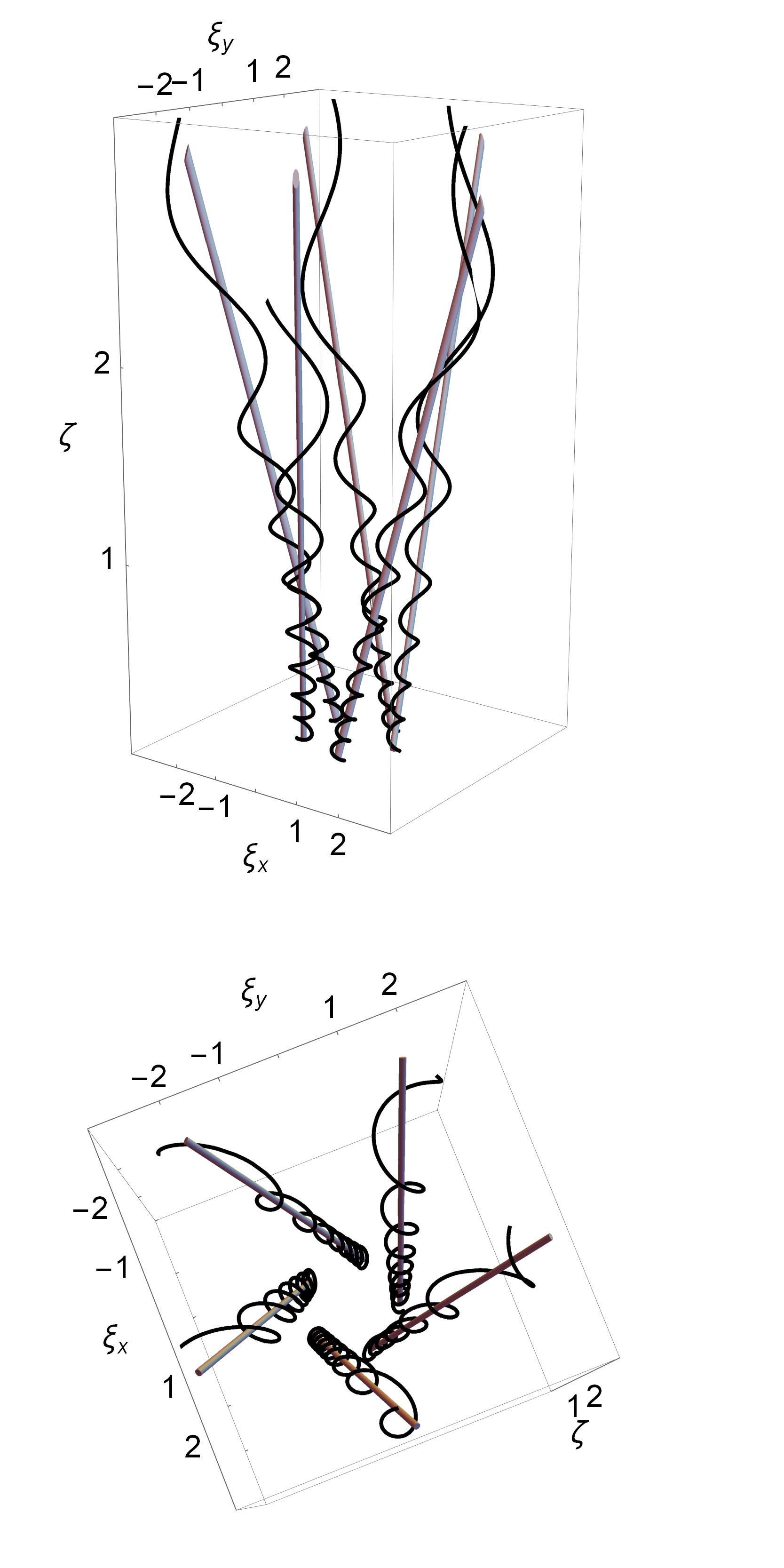}
\end{center}
\vspace{-4.5ex}
\caption{Same as in Fig. \ref{traj3}, but for five particles and for the beam of Fig. \ref{xyhole5}.}
\label{traj5}
\end{figure}

Fig. \ref{traj5} demonstrates the same phenomenon as Fig. \ref{traj3} for five particles placed in the five-hole beam of Figs. \ref{xyhole5}, \ref{xzhole5} and \ref{xyphase5}. For more complex beams, with a larger number of potential valleys, the issue becomes more challenging, owing to the very complicated arrangement of valleys and hills, which can result in jumping of particles between the tubes. The probability of such phenomenon to occur was established numerically for complex knotted vortex lines \cite{trknot}.

As has already been mentioned, the drawings representing the computed particle trajectories should be viewed as illustrative only, prepared for $\gamma$ of order of $10^{-2}$ chosen for clear visualization. A simple scaling argument leads to the conclusion that the identical effect will be achieved for smaller values of $\gamma$. However, in this case, the transported particles should be precisely placed in the potential valleys, and strongly cooled (e.g $\sim \mu \mathrm{K}$), which implies their slow motion and the necessity of determining the trajectories for very long (from the point of view of the efficiency of numerical calculations) time. 

It should also be noted that there remains at our disposal another parameter ($\beta$) that can be used to rescale the radial size of the beam structures if needed.

\begin{figure}[h!]
\begin{center}
\includegraphics[width=0.5\textwidth,angle=0]{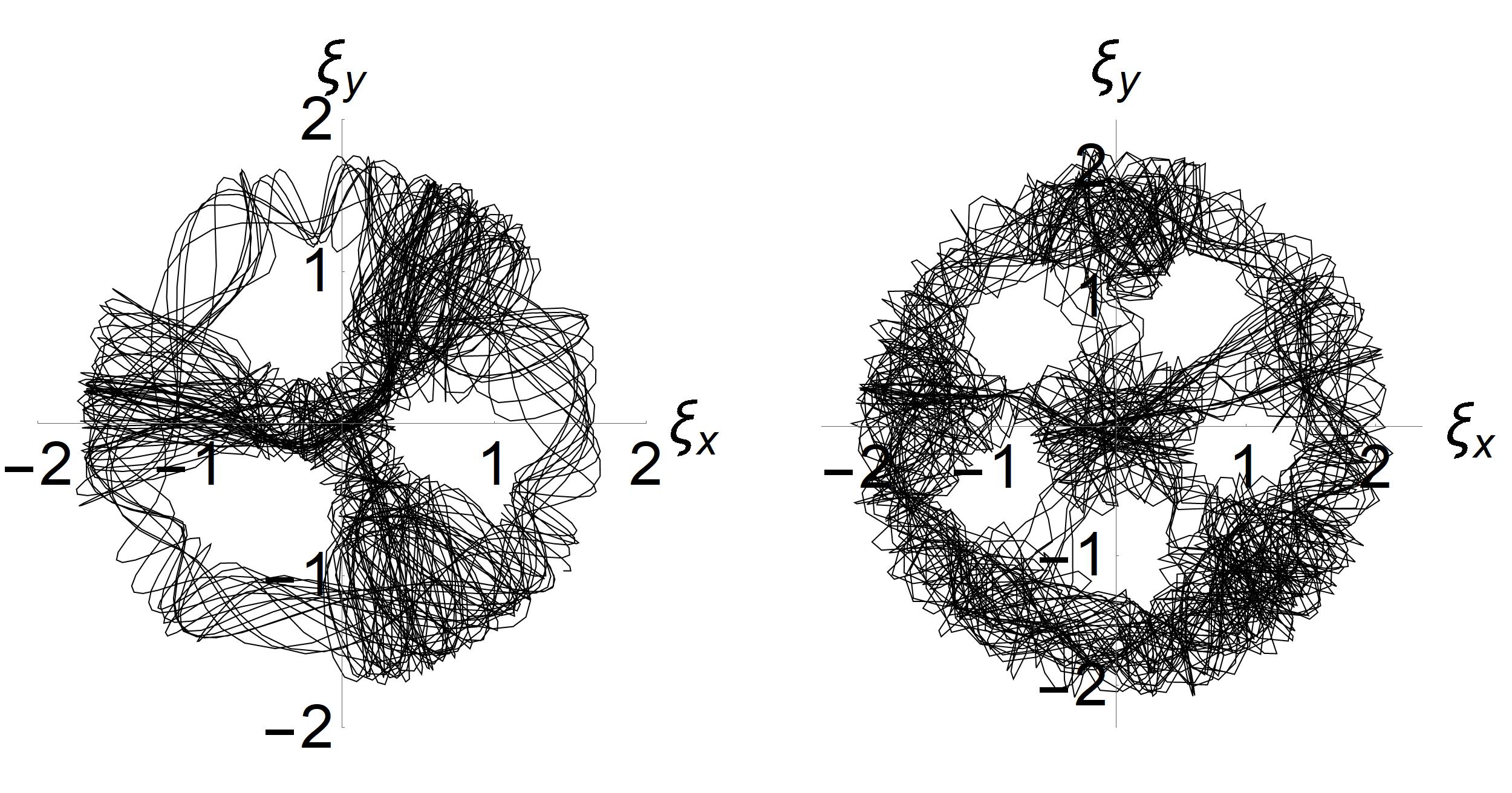}
\end{center}
\vspace{-4.5ex}
\caption{The trajctories of the positive-polarizability particle placed in the beams with $n=3$ and $n=5$ projected onto the plane $\zeta=\mathrm{const}$.}
\label{holes35}
\end{figure}

As we know, if a particle with $\alpha>0$ is inserted into a beam of this type, regions of low irradiance should exert a repulsive effect on it. As a result, there ought to remain characteristic holes in the chaotic trajectory of such a particle, which cannot be penetrated. This situation is presented in Fig. \ref{holes35} for $n=3$ and $n=5$ in the form of a projection of the calculated trajectories on the plane $\zeta=\mathrm{const}$. A three-dimensional drawing would be unreadable for obvious reasons. In line with the previous discussion, to weaken the effect of scattering force on the trajectory it was necessary to reduce the size of the spheres to $8\, \mathrm{nm}$.

\begin{figure}[h!]
\begin{center}
\includegraphics[width=0.3\textwidth,angle=0]{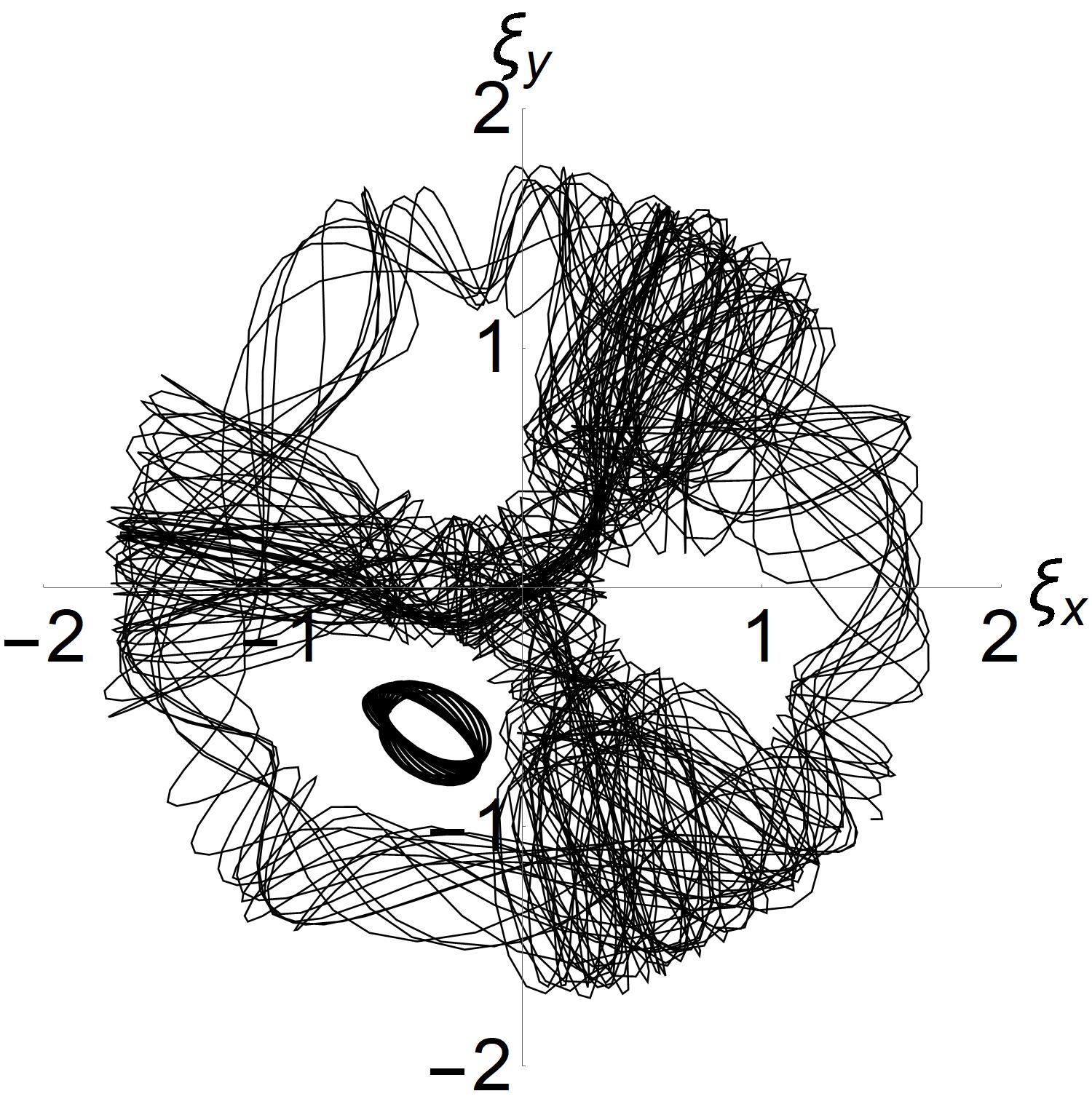}
\end{center}
\vspace{-4.5ex}
\caption{The trajectories pf two particles placed simultaneously in the beam of Fig. \ref{xyhole3}. The small trajectory corresponds to the particle of negative polarizability, and the large one to that of positive polarizability.}
\label{in3}
\end{figure}

The last figure of this section, i.e \ref{in3}, illustrates the trajectories of two particles: one with $\alpha<0$ and the other with $\alpha>0$ placed simultaneously in a beam with $n=3$. As can be observed, the former is moving inside the hole created by the trajectory of the latter.

This figure highlights the extremely chaotic nature of the positive-polarizable particle trajectories resulting from the eminently nonlinear form of the equations of motion in a very complex potential with many mninima, maxima and saddle points, versus fairly regular trajectories of negatively polarizable particles (the potential inside low irradiance tubes is quite smooth).

\section{Concluding remarks}\label{sum}
 
In conclusion, it should be stressed that the choice of a suitable Gaussian beam prefactor gives the possibility of designing beams that exhibit low-intensity tubes. In Section \ref{descr}, the prefactor was chosen in the form of a polynomial corresponding to the superposition of two Gaussian beams: a fundamental one, and one that exhibits a vortex of the $n$th degree on the propagation axis. Due to the interference, this vortex gets splitted into $n$ vortices located symmetrically on the circle, resulting in the appearance of the mentioned black wormholes. They can constitute independent lines for guiding particles.
The results of numerically performed calculations, presented in Section \ref{gpar}, show that particles with negative polarizability, neglecting some transverse oscillations, do indeed move along trajectories determined by lines of vanishing wave intensity. The influence of the scattering force on the motion has been estimated in the case of dielectric nanospheres and it has not proved to exert a significant effect on them.

Trajectories calculated for positively polarizable particles show the opposite character: they are of very chaotic nature but avoid the mentioned areas. In this case, the scattering force should be accounted for, since its impact on trajectories is significant. 

\begin{figure}[h]
\begin{center}
\includegraphics[width=0.45\textwidth,angle=0]{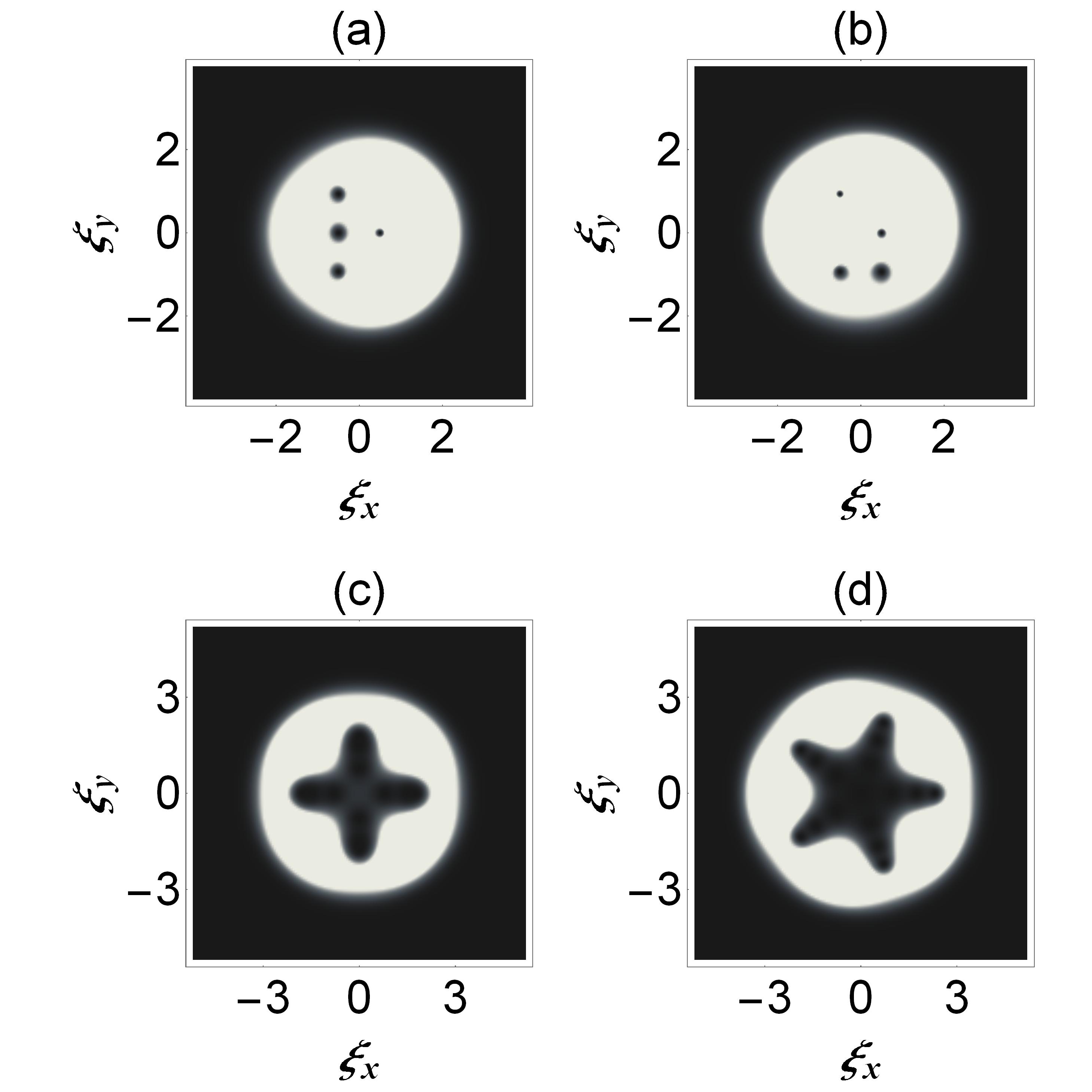}
\end{center}
\vspace{-4.5ex}
\caption{Cross sections of the beams exhibiting some special patterns: a) and b) letters `r' and `v' of the Braille alphabet, c) a cross, d) a star. }
\label{braille}
\end{figure}

Finally, it can be added that by choosing other polynomial prefactors, Gaussian beams can be obtained with various irradiance holes, designed as required. Figure~\ref{braille} shows, as some sort of curiosity, a transverse cross-section of several Gaussian beams, in which the areas of zero irradiance have been designed by choosing the appropriate polynomial prefactors.

The first two beams have hole distributions corresponding to the letters `r' and `v' from the Braille alphabet. They are generated using the superposition of five Gaussian beams of vorticity $0,1,2,3$ and $4$ (the parameter $\beta$ is maintained below which enables the pattern to be easily resized):
\begin{subequations}\label{stacr}
\begin{align}
&\tilde{\psi}_{(a)}(s)=\left( (\beta s+a)^2+b^2\right)\left((\beta s)^2-b^2\right),\label{braqr}\\
&\tilde{\psi}_{(b)}(s)=\left( (\beta s+a)^2+b^2\right)\left(\beta s -a\right)\left(\beta s -a+ib\right),\label{braqv}
\end{align}
\end{subequations}
where $a$ and $b$ are certain real constants (fixed here to be $a=0.75$ and $b=1.4$).

In order to generate the patterns representing a cross or a star more components are needed. They can be obtained correspondingly with  \begin{subequations}\label{braq}
\begin{align}
&\tilde{\psi}_{(c)}(s)=\left( (\beta s)^4-50\right)\left( (\beta s)^4-25\right)\left( (\beta s)^4-2\right),\label{crossa}\\
&\tilde{\psi}_{(d)}(s)=\left( (\beta s)^5-500\right)\left( (\beta s)^5-120\right)\left( (\beta s)^5-6\right)\beta s.\label{stara}
\end{align}
\end{subequations}
These are high-order polynomials, which means that high-vorticity beams are required to interfere. In the first case one needs four beams of vorticity $0, 4, 8$ and $12$ with appropriate relative intensities, and in the second one $1, 6, 11$ and $16$. 

\end{document}